\documentclass[3p, number, sort&compress, fleqn,authoryear]{elsarticle}
\usepackage{graphicx}
\usepackage{color}
\usepackage{amssymb}
\usepackage{amsmath}

\begin{document}
\pagestyle{empty}

\title{Elasto-capillary collapse of floating structures \\
  Non-linear response of elastic structures under capillary forces }

\author[Liege]{N. Adami\corref{cor1}} 
\ead{N. Adami@ulg.ac.be}
\author[Paris]{A. Delbos}
\author[Paris]{B. Roman}
\author[Paris]{J. Bico}
\author[Liege]{H. Caps}

\address[Liege]{GRASP -- Universit\'{e} de Li\`{e}ge, physics departement B5, 17 all\'ee du 6 Ao\^ut B4000 Li\`{e}ge, Belgium}
\address[Paris]{Physique et M\'ecanique des Milieux H\'et\'erog\`enes, ESPCI, Paris 6, Paris 7, UMR CNRS 7636, 10 rue Vauquelin, 75231 Paris Cedex 05, France}
\cortext[cor1]{Corresponding author}

\date{\today}

\begin{abstract}
Flexible rings and rectangle structures floating at the surface of water are prone to deflect under the action of surface pressure induced by the addition of surfactant molecules on the bath.
While the frames of rectangles bend inward or outward for any surface pressure difference, circles are only deformed by compression beyond a critical buckling load. 
However, compressed frames also undergo a secondary buckling instability leading to a rhoboidal shape.
Following the pioneering works of \cite{Hu} and \cite{Zell}, we describe both experimentally and theoretically the different elasto-capillary deflection and buckling modes  as a function of the material parameters.
In particular we show how this original fluid structure interaction may be used to probe the adsorption of surfactant molecules at liquid interfaces.

 \end{abstract}



\maketitle

\section{Introduction}

REORIENT the intro :\\
-lots of attention on elasto-capillary problems lately \\
-on floating structures the interest was in measuring surface tension, the focus was on small deformation
where a linear deflection law was observed.\\
- here we show how non-linear effects and the ultimate fate of structures under elasto-capillary loading for simple geometry :  instability (actually known in the context of pressurized pipes), self-contacts.
This gives an estimate for the domain of validity for the method proposed in \cite{Zell} and in \cite{Hu}.
We also discuss the relevance of different situations for the measurement of surface tension. \\

NEW in this article :\\
- experiment with a ring, with the buckling threshold.\\
- experiments on inflating the rectangle (\cite{Zell} only deflated shape, but linear things are symmetric)\\
- correct the error in computation in \cite{Zell}.\\
- showing the non-linear effects : the buckling of the rectangle.

It has been known for ages that solid materials can deform under the influence of external constraints \citep{Landau}. Several kinds of behaviors can be found following the magnitude of the constraint and the physical features of the material. The theory of elasticity refers to situations for which the deformations encountered by the material are reversible once the constraint is released \citep{Landau}. Early experiments, like the Euler's thin sheet test, show that the direction of the constraint induces different kinds of deformations. A traction constraint leads to a stretching of the sheet, while a compression with the same magnitude leads to a transverse bending of the sheet.
 In fact, it can be shown by simple energetic considerations that it is easier for the sheet to bend rather than to stretch under the influence of compression. The question sustaining the present study is the following : is it possible to emphasize universal behaviors of elastic objects submitted to the surface pressure of a surfactant layer?  \\
Recent experiments have shown that elastic objects can deform under the influence of fluid system, by the mean of surface tension constraints \citep{Zell, Hu, Choidi,Pineirua, Py, Py_2, Benoit, Jose, Boudaoud, Cambau, Hure, de Langre, Neukirch, Grotberg,Yang}. This competition between elastic and interfacial energies is in the center of numerous studies \citep{de Langre, Jose}. The present study aims to explore the behaviors of soft elastic objects (say $E\sim 0.1$ to $1$ MPa, $E$ being the Young modulus of the material) under the influence of a surfactant-loaded interface surface pressure. We first describe the experimental setup and materials used. We then describe the experimental results in terms of simple energetic calculations. The last part of the paper is devoted to a more formal description of the experiments.

\section{Experimental setup and materials}
\begin{figure}[htbp]
\begin{center}
\includegraphics[height=4cm, width=7cm]{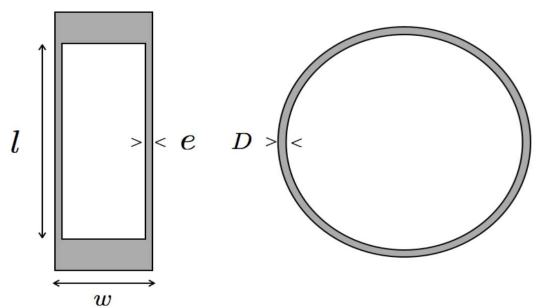}


\caption{Geometrical features of the rings and rectangles used in the experiments (view from top). }
\label{shape}
\end{center}
\end{figure}

The shape of the objects used for the present study are presented on Fig. \ref{shape}. We essentially focused on two kinds of object, being circular rings with circular section, and rectangles with square section. They are created from different types of vinylpolysiloxane (Zhermack Elite Double) which produce four types of elastic polymers characterized by Young moduli $E=0.32$ MPa, $0.66$ MPa, $0.77$ MPa, and $1.332$ Mpa, and respective densities $\rho$ of $1023$, $1140$, $1153$ and $1169$ Kg/m$^3$. Rings are produced by injection of liquid polymers in capillary tubes. The solid polymer rod obtained is then closed to induce a circular geometry. Rectangles are obtained by injecting the liquid polymer into designed molds. They are made so that only the lateral edges are sensitive to surface pressure variations. As exposed on Fig.\ref{shape} the top and bottom edges are thicker than the lateral ones, what ensures that only the lateral edges bend under surface pressure constraints. The section of the lateral arms being square, their quadratic moment is given by $I_0=e^4/12$, $e$ being both the width of the arm and the thickness of the rectangle (\emph{i.e.} the depth of the mold).  The quadratic moment linked to the rings is given by $I_0=\pi D^4/64$, $D$ being the diameter of the section. $D$ lies between $6\times 10^{-4}$ m and $1.4\times 10^{-3}$ m. The thickness $e$ of the rectangular object is fixed to 500 $\mu$m or 800 $\mu$m. The dimensions of the rectangles are $30\times12$ mm. The length of the lateral edge $l$ is fixed to $24$ mm. \\

\begin{figure}[htbp]
\begin{center}
\includegraphics[height=6.5cm, width=9.5cm]{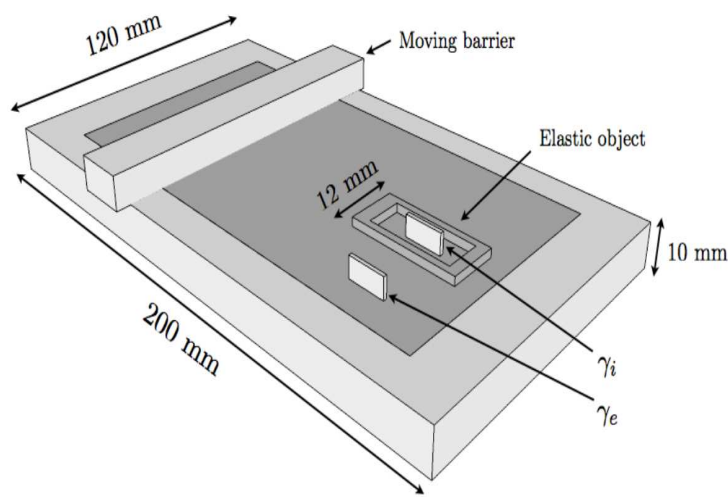}
\caption{Sketch of the experimental setup.}
\label{langmuir}
\end{center}
\end{figure}

The simplest experiment that can be made with those rings and rectangles is to lay them on a free water interface, and then to put a drop of soapy water on the outer interface. The addition of surfactant creates a surface pressure, which compresses the object. This is equivalent to a decrease of the external surface tension, leading to a force which tend to collapse the object on itself. The effects of those constraints are shown on Figs. \ref{Collapse_ring} and \ref{Collapse_rect}. One can see that the shape of the initially circular rings evolves toward a peanut shape, and that rectangles inflate when surfactant is added [PAS CLAIR POURQUOI ICI L'EFFET EST INVERSE : ON MET LE SURFACTANT DEDANS ?]. The surface pressure of a surfactant-loaded interface is expressed as :   

\begin{equation}\label{surface_pressure}
\Pi(\gamma)=\gamma_0-\gamma(\Gamma),
\end{equation}

\noindent with $\gamma_0$ the surface tension of pure water, and $\Gamma$ the surface surfactant concentration. For the needs of our study, this quantity has to be controlled, in order to relate the magnitude of the constraint on the shape adopted by the objects. To do so, a Langmuir tank is filled with millipor water. Elastic objects are deposed on the interface, as illustrated on Fig. \ref{langmuir}. Two Wilhelmy plates are used to measure the surface tension of the interface both inside and outside the object simultaneously (see Fig. \ref{langmuir}). A small quantity of surfactant molecules (ranging from 5 to 50 $\mu l$) is deposed on the outer interface. The inner interface can either be left free, or loaded with surfactant molecules as well (see next section for further details). The area of the outer interface is progressively reduced by moving the barrier of the Langmuir tank (see Fig. \ref{langmuir}), in order to increase the surface concentration in surfactant molecules outside the object, as well as the surface pressure (\ref{surface_pressure}). Since $\gamma_0$ is the same for both the inside and outside of the object, the surface pressure difference reduces to the surface tension difference, say $\Delta\Pi=\Delta\gamma=\gamma_i-\gamma_e$. To avoid diffusion of surfactant molecules from outside to inside, we used a fluorinated surfactant (CF12), which is characterized by a low bulk diffusivity. We can then assume that there is no surfactant molecule exchange between outside and inside. This can be checked by following the evolution of the inside surface tension in time when surfactant is deposed outside the object. Those observations showed that it takes about an hour for both surface tension to be equal with CF12, while it takes only a few minutes with typical commercial surfactant. Since the typical time needed to make one complete interface compression is several minutes, surfactant diffusion from outside to inside can be neglected.\\
 Pictures of the objects are taken by a camera for each value of the surface pressure difference, allowing to follow the evolution of their geometrical features (area, curvature...) with the surface pressure difference. The typical spatial resolution of the setup is $10$ $\mu m$.

\section{Experimental results}

\begin{figure}[htbp]
\begin{center}
\includegraphics[width=9cm, height=7cm]{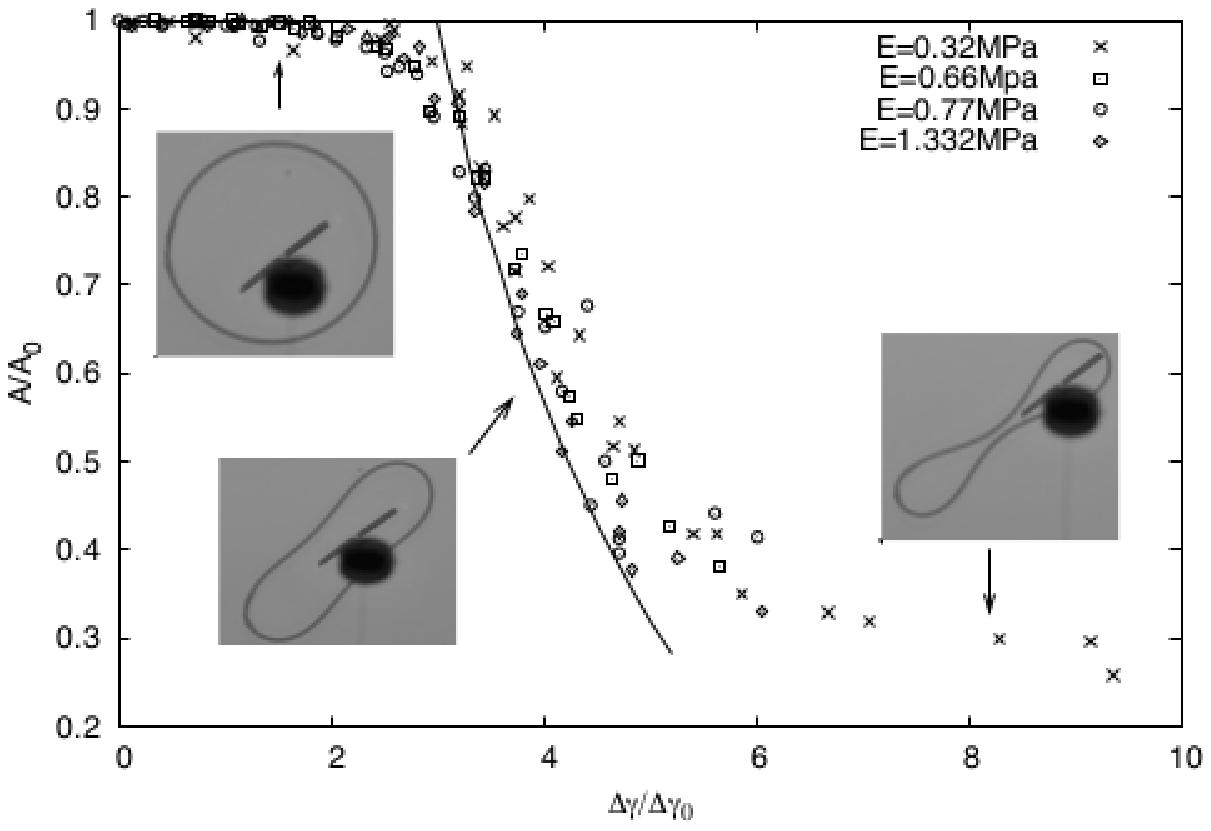} 
[REFAIRE LA FIGURE EN INCLUANT LE CALCUL AVEC AUTO-CONTACT.\\
J'AI MIS UN FICHIER SUR DROPBOX AVEC LES RESULTATS D'UN CALCUL QU'iL FAUDRAIT INCLURE ]
\caption{Evolution of the normalized area as a function of the normalized surface tension difference. Curves corresponding to different rings collapse on a master curve after renormalization. The solid line is a theoretical prediction of this collapse (see text for further details). }
\label{Collapse_ring}
\end{center}
\end{figure}

\begin{figure}[htbp]
\begin{center}
\includegraphics[width=9cm, height=7cm]{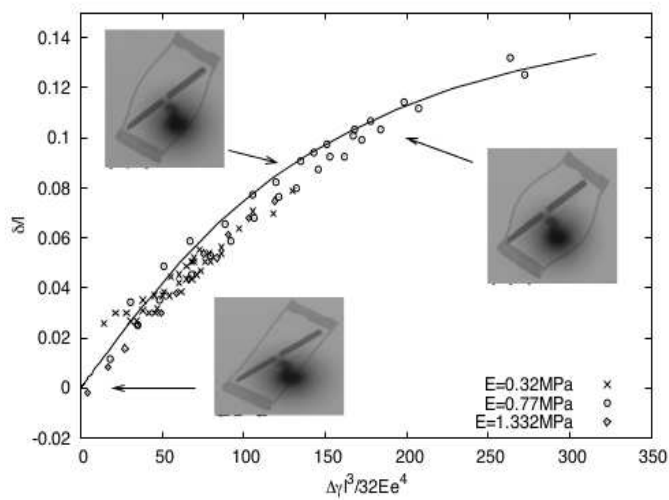}

[RAJOUTER DES POINTS SOUS COMPRESSION : LE FLAMBAGE]

\caption{Evolution of the normalized $\delta$ as a function of the normalized surface tension difference. The solid line is a theoretical prediction of this collapse. }
\label{Collapse_rect}
\end{center}
\end{figure}

As mentioned in the previous section, a decrease of the area linked to a surfactant-loaded fluid interface leads to an increase of the corresponding surface pressure (\ref{surface_pressure}). These increases in surface pressure lead to modifications in the natural shape adopted by the object when laid on a pure water interface. Figure \ref{Collapse_ring} and \ref{Collapse_rect} show several states of the compression, the most deformed corresponding to the highest surface pressure difference, and the undeformed corresponding to a weak value of the surface pressure. Experiments involving rings are performed by adding surfactant on the outside interface, so  that the surface pressure difference is increased as the area is decreased. In opposition, an amount of surfactant is added in the inner interface of the rectangles before reducing the area, what implies that the surface pressure they experience is decreased as the surface is decreased. This feature does not affect the measurements, since only the value of the surface pressure influences the shape taken by the rectangle. If surfactant was added only outside of them, one would see concave deformations of the rectangle, symmetrical to the convex ones shown on Fig. \ref{Collapse_rect}. Experiments involving free-surfactant inner interfaces for rectangles have been performed by \citep{Zell}, and have revealed rectangular object to be well-suited to realize surface tension measurements \citep{Zell, Hu}.\\
Simultaneous use of Wilhelmy plates and pictures allows to follow the evolution of the shape of the objects as a function of the surface pressure difference.

\emph{Circular annulus}

The behavior of the area sustained by rings can be easily followed from images analysis as a function of the surface tension difference imposed to the ring. Figure \ref{Collapse_ring} is a plot of this area normalized by the initial area of the ring (before surfactant is added at the interface). The surface tension has been normalized by the typical surface tension difference $\Delta\gamma_0\sim EI_0/R_0^3$ (see next section for details) with $R_0$ the radius of the ring, obtained from (\ref{elasto_cap}). 

This Figure shows the collapse obtained by the normalization. Those curves correspond to rings characterized by different values of the section, perimeter, and Young modulus. 
One can also see that a threshold must be reached in order that the area sustained by the rings starts to exhibit considerable decay. This critical value is $\Delta\gamma_0/ EI_0/R_0^3 \sim  3$. Before this critical surface tension, the annulus remains circular.

Above a given value of the load parameter [HOW MUCH?], self contact is observed, and the structure becomes stiffer (smaller rate of area change with loading). The adimensioned area decreases to 0 as the adimensioned surface tension difference increases. However, highly collapsed states of the ring could not be reached experimentally, because the largest value of the surface tension difference is fixed by the dynamical properties of the surfactant (i.e. its maximal surface concentration).\\

\emph{Rectangles}

In the case of rectangles, the deflection maximal lateral edge deflection $\delta$ (as illustrated on Fig. \ref{deform}) of the the edge is monitored. Contrary to the case of the
circular ring, deflection are observed as soon as load is applied.
The different experimental curve of the normalized deflection versus load are presented in Figure \ref{Collapse_rect}. $\delta$ has been normalized by $l$, whereas surface tension was 
adimensionalized by $32 Ee^4/ l^3$ (see next section for justification) [ON SE DEMANDE POURQUOI FACTEUR 32 ICI] leading to a collapse of the curves linked to different rectangles onto a single master curve. 
 [ICI RAJOUTER LES OBSERVATiONS SOUS COMPRESSION, FLAMBAGE, ETC]
 
 \section{Calculations and modeling}
 
 We now turn to theoretical description of the phenomena observed, using scaling laws, some exact results, and numerical integration of non-linear beam equations.
These results are compared to our experiments, but also to the existing descriptions.
In the case of the rectangle, we show that the models in \cite{Zell} and \cite{Hu} 
is only valid for very weak loads and small aspect ratio $w/l$ of the rectangle (which is relevant in their experiments).
Indeed, an extra term should be included in the description, which is in particular responsible for a buckling instability. We give a better analytical solution and characterize the buckling instability.
  
 \subsection{The circular annulus}

The floating circular annulus loaded by a surface tension difference is exactly equivalent to a pressurized tube
which can undergo buckling instability, an old problem in the mechanics of structures  (see for example early work by \cite{fairbairn1858}).

If surfactant is poured inside the ring ($\gamma_i < \gamma_e$), no change of shape is expected apart from negligible stretching of the annulus. However  
under external pressure (in our case when $\gamma_i > \gamma_e$) the axisymmetric (circular) solution always exists, but becomes unstable above a buckling threshold. 
 

This critical condition for buckling can be estimated through a scaling argument.
The energy linked to the bending of a flexible object is expressed as $\varepsilon\sim EI_0L/R^2$, with $R$ the radius of curvature linked to the bending, and $L$ the typical length along which the bending occurs. In the case of the collapse of a ring under a surface pressure difference, the bending of the ring compensate the gain of interfacial energy due to the decrease of the inner area. The typical radius of curvature during strong buckling can be estimated as the radius of the ring, as well as the typical length on which the bending occurs. The bending energy of a compressed ring is  then expressed as $E I_0/R_0$, $R_0$ being the natural radius of the ring. The corresponding gain in interfacial energy is given by $\Delta\gamma R_0^2$.  If one equals those energies, a characteristic surface tension difference can be brought out: 
\begin{equation}
\Delta\gamma_0 =  \frac{E I_0}{R_0^3}.
\end{equation}
This typical value has been used to normalize the surface pressure difference (see Fig. \ref{Collapse_ring}).
We note that the same argument leads, for a given surface tension difference, to a typical size of the system
\begin{equation}\label{elasto_cap}
L_\gamma\sim \Big(\frac{E I_0}{\Delta\gamma}\Big)^{1/3}.
\end{equation}
above which surface tension effects will lead to large deflections.
\noindent This typical length, fixed by both geometrical and physical features of the system, is an
 elasto-capillary length for floating bidimensionnal systems. The elasto-capillary length $\sqrt{Eh^3/\gamma}$ appears to be the relevant parameter in numerous experiments involving both capillarity and thin sheet elasticity \citep{Zell, Hu, Choidi,Pineirua, Py, Py_2, Benoit, Jose, Boudaoud, Cambau, Hure, de Langre, Neukirch, Grotberg,Yang}. \\

A precise computation of the equilibrium shape is obtained through non-linear beam inextensible beam theory.
If the external forces density per unit length applied  is described by $\mathbf{K}$, then the shape adopted by the object is described by the following set of equations \citep{Landau}: 
\begin{eqnarray}
\frac{d\mathbf R}{ds}&=&-\mathbf{K}, \label{F} \\
E I_0\frac{d^2\theta(s)}{ds^2}&= & \left( \mathbf{R}\times\mathbf{t} \right).\mathbf{e}_y, \label{Elastica} 
\end{eqnarray}
\noindent with $\mathbf{R}$ being the internal forces into the object, $\mathbf{t}$ the unit tangent vector to the curve described by $s$, and $\theta$ the angle defined by $\mathbf{t}$ and $\mathbf{z}$ (see Fig. \ref{deform}). Equation (\ref{Elastica}) is the well-known Euler's Elastica, which links the shape adopted by an elastic beam to external constraints. If all the relevant quantities are known, those equations can be solved numerically. Adimensionalization of those equation lead to natural appearance of (\ref{elasto_cap}) as a control parameter of the system. The external forces density $\mathbf{K}$ per unit length is simply expressed by $\Delta\gamma\mathbf{n}$, with $\mathbf{n}$ the normal unit vector linked to the shape of the object (see Fig. \ref{deform}). The boundary conditions linked to rings are $\theta(s=0)=-\pi/2$, $\theta(s=l/4)=0$, $d\theta/ds(s=0)=0$, $d\theta/ds(s=l/4)=0$, with $l$ being the perimeter of the ring. We assume that the shape of the ring is symmetric. The condition on the first derivative of $\theta$ ensures the continuity of the shape in $s=0$ and $s=l/4$. Taking those boundary conditions into account allows a numerical investigation of the shape adopted by the ring under a given surface tension difference. It also allows to get a numerical expression for the curves presented on Fig. \ref{Collapse_ring}. The solid line represents the numerical expression of the adimensioned area versus adimensioned surface tension difference obtained by solving (\ref{F}) and (\ref{Elastica}). Both area collapse threshold and decay are well reproduced.\\
In fact the buckling threshold $\Delta \gamma / \Delta\gamma_0 = ???$ is given by linear stability analysis of these equations. [TROUVER L'ARTICLE QUI PRESENTE CA]

[TO BE WRITTEN CORRECTLY:]
We can solve this up to the point where self-contacting occurs. Then we chance the numerics to allow for a force to be applied at the frictionless contact (the ring is assumed to not adhere on itself).
[MORE DETAILS HERE ABOUT BOUNDARY CONDITIONS?]
We find that the curvature at contact decreases with increasing loading, and becomes negative, and the 
ring would self-intersect agin. This means that when the curvature vanishes, another type of solution appears, 
with an extended flat contact. These solutions are found as the sum of a straight segment plus an elastica with the remaining length, starting with a zero curvature [MORE ABOUT BOUNDARY CONDITIONS]. 
When we list the unknowns, we see that this last problem only involves one length and scale-free boundary conditions, so that we can find a self-similar solution which is just scaled down. This is similar to the solution exhibited in \cite{flaherty} for buckled pressurized tube, or \cite{Mora} in the case of an elastic strip maintained in contact with itself by a soap film.
We therefore expect larger loading to lead to similar buckle, but scaled down, so that its characteristic size is proportionnal to $L_\gamma$, so that the area goes like $L_gamma^2$ and finally, we expect the area to decrease like $\Delta\gamma^{-2/3}.$ [NEW FIGURE TO SHOW THIS? OR TRY TO INCLUDE IN FIGURE].
The prefactor is actually ????. 
There will not be more change of shape, than this self flattening into a double segment whose edges are rounded with a decreasing radius of curvature (leading to material damage at some point).
[DISCUSS THE CASE OF ADHESION ?]

\subsection{Rectangles}

\begin{figure}[htbp]
\begin{center}
\includegraphics[width=9cm, height=4.5cm]{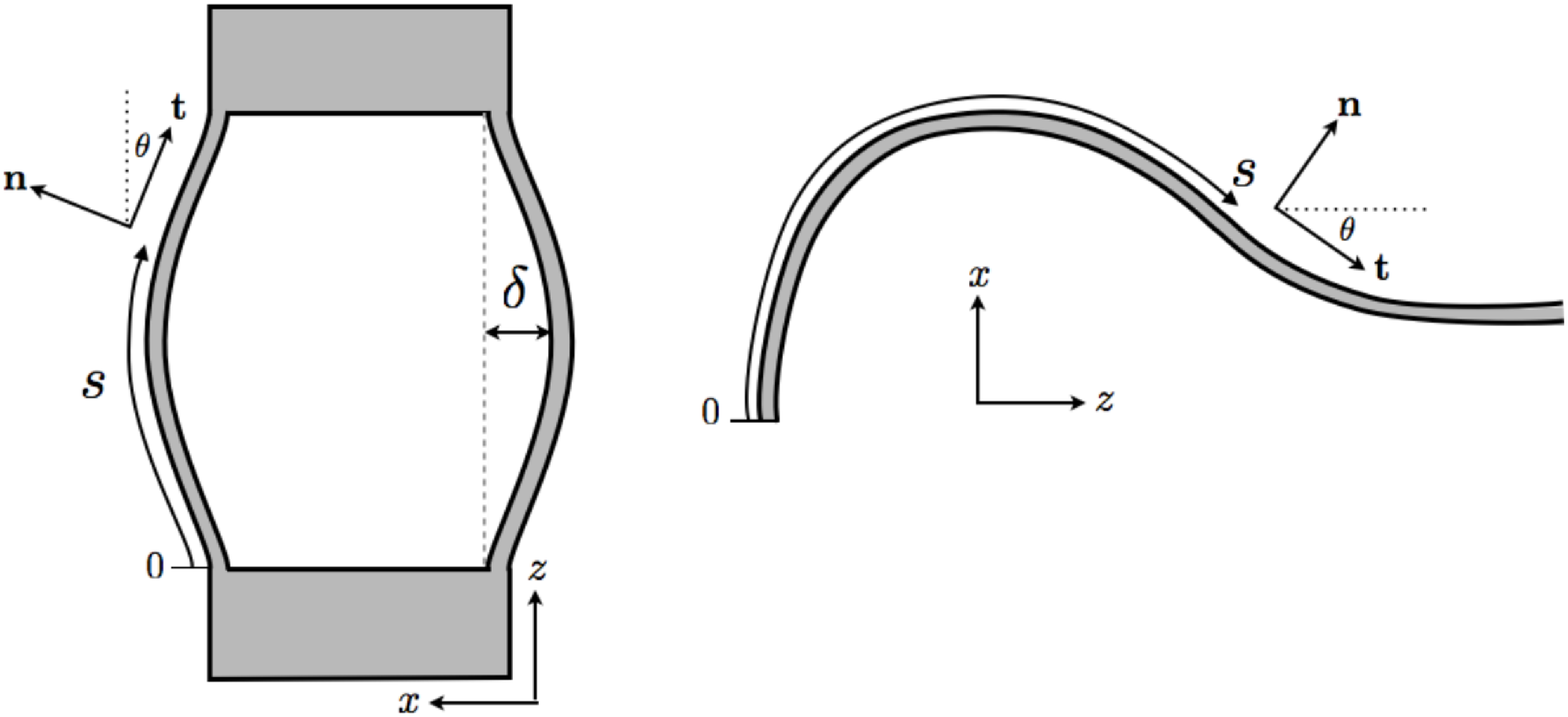}

[ATTENTION CE SCHEMA EST BIZARRE ON DIRAIT QUE LE FILAMENT COMMENCE PARALLEMENT A $e_x$ A DROIT ET PERPENDICULAIREMENT A $e_x$ A GAUCHE.] 
[PLACER LES COINS A B C D SUR LE RECTANGLE]
\caption{Scheme of the deformed object under the surface pressure constraint (view from top). Only one quarter of the deformed ring is drawn.}
\label{deform}
\end{center}
\end{figure}

We first study the case of rectangles (Fig.\ref{deform}) using scaling arguments.

A first approach is to consider the linear response of the system to very small load $\Delta \gamma$ (to be quantified later). In this regime, the deflection $\delta$ is proportional to the load, and is also very small $\delta \ll l$. In terms of energy, we therefore only keep quadratic terms in $\Delta\gamma$ and $\delta$.

The typical radius of curvature linked to the bending can then be expressed as $R\sim l^2/\delta$, with $\delta$ the maximum deflection of the lateral edge. The bending energy can be estimated as $\varepsilon_B\sim EI(\delta/l^2)^2l$. The gain in interfacial energy is estimated by $\varepsilon_\gamma\sim \Delta\gamma\delta l$, where $\delta l$ represents the area under the deflected edge.
We note that another consequence of lateral bending is the end shortening by a distance of the order of $\delta^2 /l$.  This corresponds to an extra interfacial energy gain $\varepsilon_w \sim \Delta\gamma w \delta^2 l$ which is of third order in $\delta,\Delta\gamma$ and therefore does not contribute to the linear response theory.

The equilibrium shape minimizes the sum of  energies $  \varepsilon_B  - \varepsilon_\gamma$ , and follows
\begin{equation}\label{delta_rect}
\delta \sim \frac{\Delta\gamma l^4}{E I_0}.
\end{equation}
\noindent
The prefactor in this expression was computed in \citep{Zell}. Assuming small slope, and assuming also only lateral pressure (and therefore neglecting axial end forces due to pressure on the width $w$), the linearized beam equation~(\ref{F},\ref{Elastica}) read
$EI d^3y/dz^4 +R=0,$ where  $dR/dz=\Delta \gamma$, and finally $d^4y/dz^4 + \Delta \gamma=0.$ 
The polynomial solution consistent with boundary conditions is therefore
\begin{equation}
x= \Delta\gamma l^4/EI_0 ( - z^2/24 +z^3/12 - z^4/24) \label{x_rep_lin}
\end{equation}
 Taking into account the expression for the quadratic momentum of a square section and the fact that
  (\ref{delta_rect}) must be satisfied for $z=l/2$, one finds an analytical expression for $\delta$, being : 
\begin{equation}\label{delta_rect_formal}
\delta =  \frac{\Delta\gamma l^4}{384EI_0} = \frac{\Delta\gamma l^4}{32Ee^4}.
\end{equation}
\noindent This expression has been successfully used to measure surface pressure from deflection of rectangular objects \citep{Zell,Hu}. In these experiments the deflection was always symmetric.
One could conversely use this expression to determine the Young Modulus of a given polymer, by analysis of the deformations induces by a known surface pressure. 

However, in our experiments we sometimes observe a sudden twisting of the rectangle which completely invalidates the measurements suggested.
The remaining of this section is devoted to characterize the non-linear response of the system, explain the buckling instability and determine the regime where  (\ref{delta_rect_formal}) can be used.

It is instructive to start with scaling argument. We still consider small slope $\delta \ll l$, but now include the surface energy $\varepsilon_w$ that was discarded. The equilibirum shape minimizes  $  \varepsilon_B  - \varepsilon_\gamma- \varepsilon_w$, and we find 
\begin{equation}\label{Buckling_nrj}
[EI_0/l^4 - \alpha  \Delta \gamma w/l^2 ] \delta \sim  \Delta \gamma,
\end{equation}
where $\alpha$ is a parameter that we don't compute here. This equation shows that the effective rigidity of the system $k=EI_0/l^4 - \alpha  \Delta \gamma w/l^2 $ is decreased by the non-linear effects considered. We also see that we recover the linear response (\ref{delta_rect}) when $\Delta \gamma \ll \Delta \gamma_r =  EI_0/(wl^2).$ Finally we see that for a critical value of  $\Delta \gamma  / \Delta \gamma_r$, the rigidity of the system vanishes and becomes negative for larger loads. This is the sign of a mechanical instability, such as buckling.

We now turn to a more rigorous version of the same argument, based on beam equations~(\ref{F},\ref{Elastica}),
where $\mathbf{K}=\Delta \gamma \mathbf{n}.$ These equations are to be applied all along the flexible sides of the rectangles. Along the rigid sides  $\theta$ is constant, but equation (\ref{F}) still holds.
for $s=0 to l$. The clamped boundary conditions are $\theta(0)=\theta(l)=0$, but the boundary conditions on the force $\mathbf{R}$ are less obvious. We consider that the systems maintain a diagonal symmetry such that the corner (A,C) of the rectangle (ABCD) lying on a diagonal are under indentical mechanical state.
We obtain two more conditions in a relation between $\mathbf{R}(0)=\mathbf{R}_A$ because symmetry imposes $\mathbf{R_C}=-\mathbf{R_A}$, but $\mathbf{R_C} = \mathbf{R_A} + \Delta \gamma\left[ \left(w+x(l) \right) \mathbf{e}_z +  z(l)\mathbf{e}_x \right]$. [CHECK SIGNS]
These ODE and the corresponding four boundary conditions can be solved using a shooting technique.

\begin{figure}[htbp]
\begin{center}
\includegraphics[width=16cm]{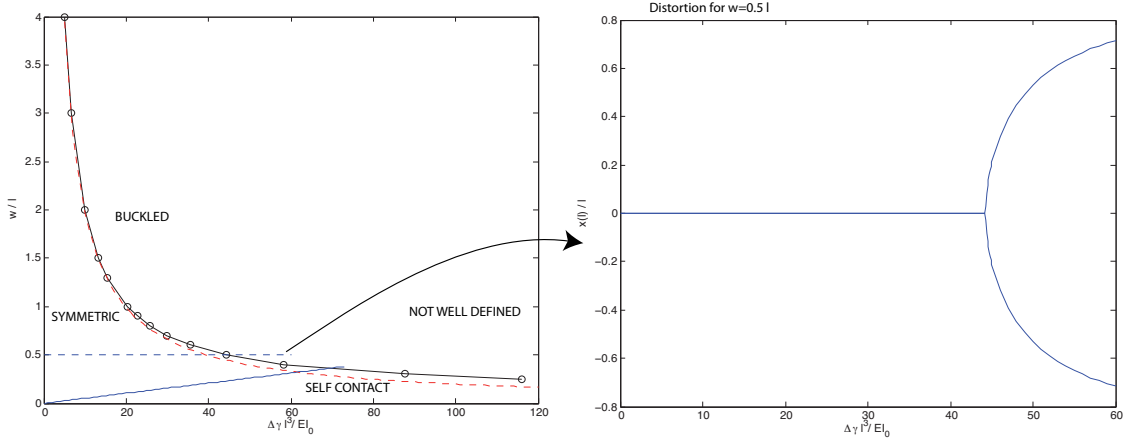}
\caption{Numerical results. LEFT : phase diagram  in the plane ($w/l$, $\Delta\gamma l^3/EI_0$). above the 
dark line, the system is buckled in the numerical integration of full equations. In red dashed line is presented the buckling threshold in the small small approximation, separating the symmetric from buckled state. 
the blue line is the prediction of self intersection. [INCLUDE NUMERICAL COMPUTATION OF THIS CONDITION. ALSO COMPUTE  SELF CONTACT IN BUCKLED REGION] .right bcukling deflection as a function of $Delta\gamma l^3/EI_0$ for a given $w/l=0.5$ corresponding to the cut in the phase diagram on the left. [WE CAN ALSO PRESENT AREA AS FUNCTION OF LOAD, OR $\delta = x(z=1/2)$]
}
\label{rect_numerique}
\end{center}
\end{figure}

[DEVELOP]Ê Comment on figure~\ref{rect_numerique}. In the numerical resolution we find a buckling instability, through a pitchfork bifurcation . A symmetry is broken. When aspect ratio $w/l$ is larger than ????, then buckling  occurs before self contact .
What values of aspect ratio allows the largest pressure measurement for a given rigidity?

Some predictions can be made using analytic solutions in the
in the small slope approximation $x/l\ll 1$. The equations~(\ref{F},\ref{Elastica}) becomes
\begin{eqnarray}
\frac{d\mathbf R}{dz}&=&- \Delta \gamma \mathbf{e}_x, \\
E I_0\frac{d^3 x(z)}{dz^3}&= & \left( \mathbf{R}\times\mathbf{e}_z \right).\mathbf{e}_y, 
\end{eqnarray}
[CHECK THE SIGN!] \\
These equations lead to 
$$ d^3x/dz^3 + ( p w /2) dx/dz + p (1-2z)/2 = 0 $$
where distances are non-dimensionalized by $l$ (for example $w$ stands for $w/l$), and $p=\Delta\gamma l^3/ EI_0$. This linear ODE leads to
$$x = (z^2-z)/w + A[\cos(  z\sqrt{pw/2} + \phi ) - \cos\phi]$$
where $\phi$ and $A$ are to be determined with boundary conditions $y'(0)=y'(1)=0$.
\begin{eqnarray}
 A &=&  -\sqrt{2/pw^3} / \sin\phi \\
 \sin(\sqrt{pw/2} +\phi) &=& -\sin\phi
 \end{eqnarray}
 We see that the solution is given by  $-\phi= \sqrt{pw/2} +\phi$, so that $\phi = -\sqrt{pw/8}$ and
 \begin{equation} x_0(z) = (z^2-z)/w +  -\sqrt{2/pw^3} / \sin( \sqrt{pw/8})  \left[\cos(  z\sqrt{pw/2}  -\sqrt{pw/8} ) - \cos \sqrt{pw/8}\right]
 \label{x_lin}
 \end{equation}
$x_0(z)$ is symmetric around $z=1/2$, so that  $x_0(1) = x_0(0)$.
Although the analytic expression is different from the linear response (\ref{x_rep_lin}), they share the same symmetry, and are very close. Indeed when $pw\to 0$, expression (\ref{x_lin}) converge towards (\ref{x_rep_lin}) and therefore towards (\ref{delta_rect_formal}), as can be seen using a Taylor expansion.
\begin{figure}[htbp]
\begin{center}
\includegraphics[width=16cm]{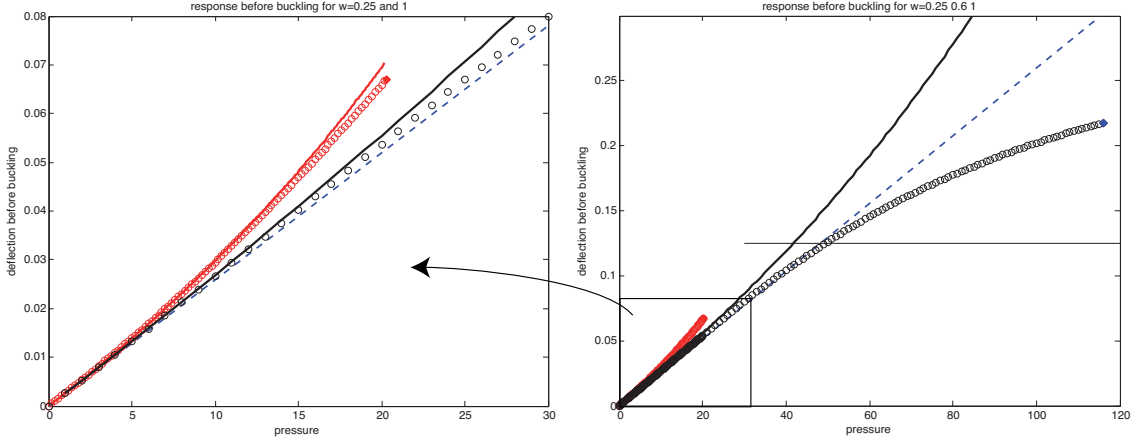}
\caption{adimensional deflection $\delta/l$ versus $\Delta\gamma l^3/EI_0$ before buckling instability. The linear response theory (dotted blue line) is compared to numerical integration (circles) and also to analytical description (continuous line) for $w/l=1$ (in red) and $w/l=0.25$ in black (value used in~\cite{Zell}). The $w/l=1$ curve stops because of buckling instability (red diamond), whereas the computation of the $w/l=0.25$ curve hits a self contact when $\delta/l=w/(2l)=0.125$ (horizontal line) before buckling (black diamond) }
\label{comp_response}
\end{center}
\end{figure}

We see on two examples with $w/l=1$ (case of a square) and $w/l=0.25$ (value extracted from pictures in~\cite{Zell}) how the linear response is very close to numerical integration of equations for vanishing $\Delta\gamma l^3/EI_0.$
The deflection predicted (see figure~\ref{comp_response}) are always larger than the  linear response theory (the scaling argument did suggest that including the pressure on the rigid wall would lead to a decrease of apparent rigidity). 

For $w/l=1$ (case of a square), this axial pressure effect is noticeable as the numerical curve deviates significantly from the linear response theory.  The calculated curve however does follow very closely the numerical results, up to a buckling instability (see zoom on left of  figure~\ref{comp_response}).
For small aspect ratio however, such as $w/l=0.25$, the theories are not very different, and in fact it seems that
the linear theory does an even better job at predicting the numerical solution of equations than the refined theory. In fact in both analytical theories we have always assumed a small slope $\delta/l\ll 1$ discarding geometrical non-linear effects. These effects tends to stiffen the structure, so that the refined theory overestimates the deflection. It turns out that both effects (the weakening due to pressure on rigid walls and the non-linear stiffening) almost compensate, so that the naive linear response theory sees an extended domain of validity.
In contrast with the case of the square $w/l=1$, buckling does not occur here, as the flexible sides of the triangle contact each-other when $\delta/l = w/(2l)=0.125.$ We did not compute the evolution of the system above this point (but in figure~\ref{comp_response} we present the response above this point, with interpenetration, in order to show the dominant geometrical stiffening effect).

We compute the bifurcation threshold by considering a perturbation $\delta x(s) \ll x_0(s)$ so that $x= x_0(s)+ \delta x(s)$. The equilibrium equation for the perturbation is simply $d^3\delta x/dz^3 + (p w /2) d\delta x/dz=0$, so that
$\delta x(s)=B (\cos(\sqrt{pw/2} z + \psi)-1)$. The boundary condition $d\delta x/dz=0$ in $z=0,1$ show that if $pw \neq 2\pi^2$, $\delta x(z)=0$. However if $pw = 2\pi^2,$ then an unstable mode exists, $\delta x(s) = 1-\cos( \pi z)$, corresponding to a buckling of the structure as observed in experiment [FIGURE?].

This condition for buckling (in the limit of small slope) is therefore $p = 2 \pi^2 /w $ , or in dimensional terms
$$
\frac{ \Delta \gamma l^2 w}{EI_0} = \frac{ \Delta \gamma }{  \Delta \gamma_r} = 2\pi^2
$$
as expected from the scaling argument.
This small slope calculation of  buckling threshold is expect to be valid when takes place for small deflections, when $w/l$ is large. This can be observed on figure~\ref{rect_numerique} where the numerical detection of the buckling in the complete numerical scheme, is presented as well as the small slope prediction.

We see that the measurement of surface tension is easy when it is not perturbed by buckling instability (occurring quickly for large aspect ratio $w/l$), or contact 
between the flexible sides of the rectangle (dangerous at small $w/l$). There is therefore an optimal aspect ratio where both phenomenon take place at the same load. According to our approximate linearized theory and buckling, it follows $w/2 =  (\Delta\gamma l^4)(384EI_0) $ and $(\Delta \gamma l^2 w)/EI_0 = 2\pi^2.$
So that $w/l= 2\pi/\sqrt{384}$  [CA DEVRAIT ETRE PROCHE DE $w/l=0.3$ D'APRES LE GRAPHE].
in the numerical integration this value is in fact ????

 \subsection{Nonlinear behavior - to be removed or included above}

The collapses presented on Figure \ref{Collapse_ring} and \ref{Collapse_rect} are obtained thanks to renormalization of the control parameters linked to the experiments, say $L_\gamma$. Simple order of magnitude calculations about the energies involved in the system lead to a clear and elegant understanding of what happens to the elastic objects when they are submitted to surface pressure constraints. More formally, their behaviors under external constraints can be described in terms of continuous media mechanics.

The same kind of numerical solving can be made for rectangles. The boundary conditions are changed into $\theta(s=0)=0$, $\theta(s=l/2)=0$, so that the geometrical constraints linked to the rectangles are taken into account. The solving of (\ref{F}) and (\ref{Elastica}) leads to the solid curve presented on Fig. \ref{Collapse_rect}. Here again, experimental collapses are well reproduced theoretically. Since (\ref{delta_rect_formal}) is linear in $\Delta\gamma$, one can wonder why both experimental and theoretical curves exhibit a curbed shape for the larger values of $L_\gamma$, the smaller value having a linear-like dependency. This comes from the fact that  (\ref{delta_rect_formal}) accounts only for small deformations, while large $L_\gamma$ values are linked to considerable deformations (see Fig. \ref{Collapse_rect}).\\

[TO BE CHECKED, AND DECIDED IF INTERESTING]

For very large inflation $\Delta\gamma \to \infty$, the shapes tends to that of an inflated inextensible string, 
and therefore the area has an asymptotic maximal value.
This value can be computed by considering that the shape is a collection of arcs of circles, whose radius needs to be determined.
In this case the shape maximizes the volume $V/l^2= 2\alpha \theta^{-1} \sin(\theta/2)+  (\theta - \sin\theta)/\theta^{2}$ if $\alpha = w/l$ the size of the non-deformable "trunk".
In our experiment [BASE SUR PHOTOS DE LA FIGURE : A VERIFIER ] $\alpha = 0.55$, so that the maximum
corresponds to $\theta = 2$, and therefore $\delta/l = (1-\cos\theta/2)/\theta \sim 0.23$.
[NICOLAS : IL SERAIT INTERESSANT DE POUSSER LE CALCUL DE LA  FIGURE 4 PLUS LOIN POUR VOIR SI CA MARCHE]

[A FAIRE : LE CALCUL DU FLAMBAGE ET DE LA DISTORTION]

\section{Conclusion}

TO BE EXPANDED !

In conclusion, we have investigated the behaviors of elastic rings and rectangles under the influence of the surface pressure of a given surfactant. Those behaviors show universal features, which are well explained by simple energetic calculation, and faithfully reproduced by formal investigations of the problem. Both numerical and experimental results emphasize the relevance of the elasto capillary length as being the control parameter of the system.\\

\section{SUPPRIME CETTE PARTIE: Nouvelle version de la theorie du rectangle pour expliquer la distortion - A VERIFIER} 

\begin{figure}[htbp]
\begin{center}
\includegraphics[width=15cm]{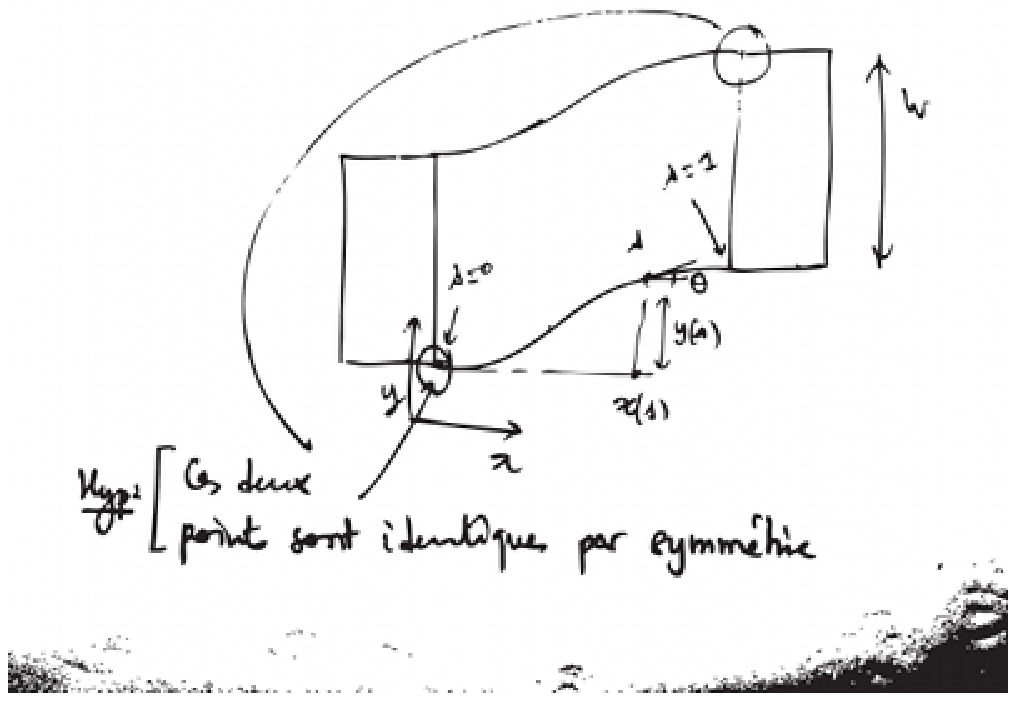}
\end{center}
\end{figure}

\subsection{conditions aux limites}
Je note $w$ la largeur du cot\'e \'epais du rectangle.
Il faut rŽsoudre les mmes Žquations (avec $p=\Delta \gamma /e$ je pense). Tout est adimenssionne par $l$ l'autre cote du rectangle.
\begin{eqnarray}
dR_x/ds &=& p \sin\theta \\
dR_y/ds &=& -p\cos\theta \\
d^2\theta/ds^2 &=& -R_y\cos\theta +R_x\sin\theta 
\end{eqnarray}
mais je pense qu'on avait faux sur les conditions aux limites: c'est plut™t
\begin{eqnarray}
\theta(0) &=&0  \\
\theta(s=1) &=&0 \\
2 R_x(0) &=& -p y(1) -p w \\
2 R_y(0) &=& p x(1) 
\end{eqnarray}
En effet on doit avoir $R(1) - pw e_z =-R(0)$ par symmetrie, et on sait que $R(1) =  R(0) + px(1) - py(1)$.
C'est cela qui change [on avait pris $\theta(0)=\theta(1/2)=0$ et $R_y(0)=px(1)/2$ et $R_x(0)=py(1)/2 $].

\subsection{resolution (linearisee) pour faibles deflections}
On peut faire une rŽsolution pour les faibles angles (linearisation) {\bf sous l'hypothese $y\ll 1$}, et donc
$$ d^3y/dx^3 + (pw/2) y' + p (1-2x)/2 = 0 $$
que l'on peut rŽsoudre 
$$y = (x^2-x)/w + A[\cos( \sqrt{pw/2}x + \phi ) - \cos\phi]$$
et il faut dŽterminer $\phi$ et $A$ avec les conditions aux limites $y'(0)=y'(1)=0$.
\begin{eqnarray}
 A &=&  \sqrt{2/pw^3} / \sin\phi \\
 \sin(\sqrt{pw/2} +\phi) &=& -\sin\phi
 \end{eqnarray}
 Deux cas possibles :
 \begin{itemize}
\item $\sqrt{pw/2}<\pi$, alors $-\phi= \sqrt{pw/2} +\phi$, donc $ \phi = \sqrt{pw/8}$ et on trouve une solution $y$ symmetrique autour de $x=1/2$, de sorte que  y(1) = y(0). Ce sont des solutions trs proches de celles qu'on pensait avoir! Normalement si $p\to 0$ on devrait retrouver les mmes, d'ailleurs. 
\item si $\sqrt{pw/2}=\pi$ alors cette derniere equation est indeterminŽe : tous les $\phi$, donc toutes les amplitudes sont possibles. Ca ressemble beaucoup au flambage.
\end{itemize}
J'en conclus que la condition de flambage (apparition de la distortion) est $$p = 2 \pi^2 /w $$ tant que cette valeur est assez petite pour qu'on reste dans les faibles amplitudes, donc pour $w$ assez grand.

\subsection{TO DO }
\begin{itemize}
\item Il faut vŽrifier si a marche avec le numŽrique (et avec les manips de Nicolas).
\item  Il faut aussi voir ce qui se passe aussi pour de plus petites valeurs de $w$ (numerique) : est-ce que a flambe toujours, ou bien est-ce qu'il y a un $w$ au dessous duquel il n'y a plus de distortion, quelle que soit la pression. Par exemple s'il y a auto-contact avant la condition de flambage.
\item Žtudier le type de transition : est-ce qu'on bien brisure de symmŽtrie? laquelle exactement? pour l'instant je  ne trouve qu'une seule branche numŽriquement.
\end{itemize}

\subsection{Young Modulus measurements (a placer soit ici en annexe, soit dans le texte)}

\begin{figure}[htbp]
\begin{center}
\includegraphics[height=7cm, width=5cm]{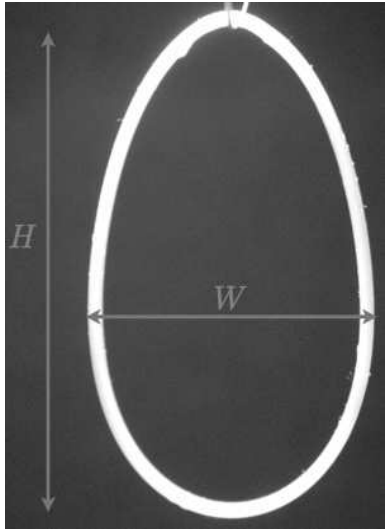}\qquad \includegraphics[height=7cm, width=9cm]{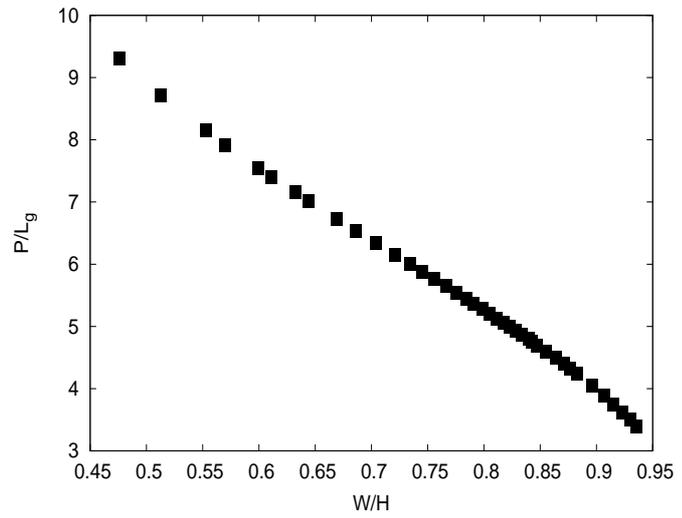}
\caption{Left : picture of a hanging ring, as considered for Young modulus measurements. Right : theoretical dependency of the ring perimeter adimensionned by the elasto-gravitary lenght $L_g$ versus the aspect ratio of the ring.}
\label{hanging_ring}
\end{center}
\end{figure}

Both order of magnitude expressions (Eq.(\ref{Buckling_nrj})) and related formal analyses (Eq.(\ref{Elastica})) used to describe the shape of soft objects due to a surface pressure load imply the Young modulus of the polymer used to build them. In order to get reliable descriptions of those objects shapes versus a given load, their Young moduli must be determined (precisely). Their measurements can be done by considering the shape adopted by rings when submitted to their weight while hanging on a nail (see Fig. \ref{hanging_ring}). For given geometrical features (\emph{i.e.} the radius $R$ and the section $S$), it can be checked that the aspect ratio $W/H$ adopted by decreases as the Young modulus increases. The bending energy can be evaluated as $\varepsilon_B\sim EI_0/R$, and the related gain of potential energy as $\varepsilon_g\sim \rho g R^2 S$, with $\rho$ the density of the polymer. Equaling $\varepsilon_B$ to $\varepsilon_g$ leads to a typical size for the rings, say the elasto-gravitary length, above which they bend under gravity, being : 
$$L_g=\Big(\frac{E I_0}{\rho g S}\Big)^{\frac{1}{3}}.$$
In order to link $L_g$ to the shape of the ring, one can solve Eq.(\ref{F}) and (\ref{Elastica}) with boundary conditions being $\theta (s=0)=0$, $\theta (s=l/2)=0$, $x(s=0)=0$ and $x(s=l/2)=0$. Adimensionnalisation of these equations lead to the appearance  of $L_g$ as unique control parameter. Figure \ref{hanging_ring} shows the ring perimeter adimensionned by $L_g$ as a function of the $W/H$. Since all quantities are known from experimental determination, it is then possible to use this curve to find the corresponding Young Modulus.  Table \ref{mesure_young} compares the Young Moduli values for different polymers, obtained by usual traction test and by the hanging ring method. Both accuracy are of the order of 10\%.

\begin{table}[htdp]\label{mesure_young}

\begin{center}
\begin{tabular}{|c|c|c|c|c|}
\hline

Material & pink & violet & dark green & light green\\
\hline
Traction test & 270 kPa & 681 kPa & 789 kPa & 1.2 Mpa\\
Hanging ring & 300 kPa & 666 kPa & 774 kPa & 1.332 Mpa\\

\hline
\end{tabular}\end{center}
\caption{Comparison of Young Moduli measurements using the usual traction test method and the hanging ring method.}

\end{table}

\section*{Acknowledgments}
NA thanks the FRS-FNRS and the Communaut\'{e} Fran\c caise de Belgique for financial support. The authors also wish to thank Gerry Boulet for emotional support during the experiments(private joke, to be removed before submission, of course).


\bibliographystyle{elsarticle-harv}

\end{document}